\documentclass[prl,amsmath,amssymb,superscriptaddress,nofootinbib,twocolumn]{revtex4}
\usepackage{amsfonts}
\usepackage{multirow}
\usepackage{makecell}
\usepackage{mathrsfs}
\usepackage{graphicx}
\usepackage{amsmath}
\usepackage{amssymb}
\usepackage{bm}
\usepackage{bbm}
\usepackage{color}
\usepackage{ulem}
\usepackage{array}
\usepackage{diagbox}
\usepackage{float}
\usepackage{orcidlink}

\newcommand{\nn}{\nonumber}

\newcommand{\beq}{\begin{equation}}
\newcommand{\eeq}{\end{equation}}
\newcommand{\bqa}{\begin{eqnarray}}
\newcommand{\eqa}{\end{eqnarray}}

\newcommand{\bseq}{\begin{subequations}}
\newcommand{\eseq}{\end{subequations}}

\makeatletter

\begin{document}

\title{ Next-to-leading-order QCD corrections to nucleon Dirac form factors
}
\author{Long-Bin Chen$^{\orcidlink{0000-0002-7647-7716}}$}
\affiliation{ School of Physics and Materials Science, Guangzhou University, Guangzhou 510006, China \vspace{0.2cm}}
\author{Wen Chen$^{\orcidlink{0000-0002-7568-2056}}$}
\affiliation{ Key Laboratory of Atomic and Subatomic Structure and Quantum Control (MOE), Guangdong Basic Research Center of Excellence for Structure and Fundamental Interactions of Matter, Institute of Quantum Matter, South China Normal University, Guangzhou 510006, China\vspace{0.2cm}}
\affiliation{Guangdong-Hong Kong Joint Laboratory of Quantum Matter, Guangdong Provincial Key Laboratory of Nuclear Science, Southern Nuclear Science Computing Center, South China Normal University, Guangzhou 510006, China\vspace{0.2cm}}
\author{Feng Feng$^{\orcidlink{0000-0003-3608-1108}}$~\footnote{Corresponding author: f.feng@outlook.com}}
\affiliation{China University of Mining and Technology, Beijing 100083, China\vspace{0.2cm}}
\affiliation{Institute of High Energy Physics, Chinese Academy of Sciences, Beijing 100049, China\vspace{0.2cm}}
\author{Siwei Hu$^{\orcidlink{0000-0001-6491-3034}}$}
\affiliation{Institute of High Energy Physics, Chinese Academy of Sciences, Beijing 100049, China\vspace{0.2cm}}
\affiliation{School of Physical Sciences,
University of Chinese Academy of Sciences, Beijing 100049, China\vspace{0.2cm}}
\author{Yu Jia$^{\orcidlink{0000-0001-8674-2790}}$~\footnote{Corresponding author: jiay@ihep.ac.cn}}
\affiliation{Institute of High Energy Physics, Chinese Academy of Sciences, Beijing 100049, China\vspace{0.2cm}}
\affiliation{School of Physical Sciences,
University of Chinese Academy of Sciences, Beijing 100049, China\vspace{0.2cm}}
\date{\today}

\begin{abstract}
The leading-order perturbative QCD (pQCD) predictions to nucleon electromagnetic form factors were first made in late 70s.
In this Letter for the first time we accomplish the calculation of the next-to-leading-order (NLO) QCD corrections to
nucleon's Dirac form factors at large momentum transfer, to the leading-twist accuracy in collinear factorization approach,
specifically within the Krankl and Manashov renormalization scheme.
The effect of NLO perturbative corrections turns out to be positive and substantial.
Taking the nucleon leading-twist light-cone-distribution amplitudes (LCDAs) determined from the recent lattice simulations as input,
we find that the state-of-the-art pQCD predictions significantly underestimate the available nucleon Dirac form factors
in both space-like and time-like domains. This nuisance indicates that some
additional soft nonfactorizable contribution might be called for to account for the measured nucleon electromagnetic form factor data
up to $Q^2\approx 30\;{\rm GeV^2}$.
\end{abstract}

\maketitle

\noindent{\color{blue}\it Introduction.}  As the composite particles made of strongly-coupled, relativistic quarks and gluons,
proton and neutron are the building block of the atomic nuclei, consequently, of our planet and
all its inhabitants, and even more, of our visible universe.
Understanding the internal structure of nucleons from quantum chromodynamics (QCD), the underlying
theory of strong interaction, is the fundamental challenge faced by the contemporary hadron and nuclear physics~\cite{Gross:2022hyw}.

Nucleon electromagnetic form factors (EMFFs), which gauge the distributions of the electric charge and magnetization inside the nucleon,
are the fundamental probes to the internal structure of nucleon~\cite{Gross:2022hyw}.
The nucleon EMFFs can be directly accessed from the $ep$ elastic scattering experiments,
which are encoded in the following nucleon transition matrix elements:
\beq
\langle P'\vert j^{\mu}_{\rm em} \vert P\rangle = \overline{N}
\left[F_1(Q^2) \gamma^\mu - F_2(Q^2) \frac{i\sigma^{\mu\nu}q_\nu}{2m_N} \right] N,
\label{nucleon:formfactor}
\eeq
with the electromagnetic current $j^\mu_{\rm em}=\sum_f e_f \bar{\psi}_f \gamma^\mu \psi_f$.
$N$, $m_N$
denote the Dirac spinor and mass of the nucleon, respectively.
$q^\mu\equiv P^\mu-P'^\mu$, and $Q^2\equiv -q^2\ge 0$ characterizes the squared momentum transfer.
The scalar functions $F_1$, $F_2$ are referred to as the (helicity-conserving) Dirac form factor,
and the (helicity-flipped) Pauli form factor, respectively.

After Hofstader's pioneering work in mid-50s~\cite{Hofstadter:1955ae,Mcallister:1956ng},
intensive experimental efforts have been continuously devoted to measuring the nucleon EMFFs $F^{p,n}_{1,2}$
over the past seven decades (for comprehensive experimental review, see \cite{Gao:2003ag,Hyde:2004gef,Perdrisat:2006hj,Punjabi:2015bba}).
In early 90s {\tt SLAC} has already measured the space-like proton EMFFs at $Q^2=31.3\;{\rm GeV}^2$~\cite{Sill:1992qw}.
In the prospective electron-ion colliders {\tt EIC} and {\tt EicC}, the momentum transfer for nucleon EMFFs may be further
extended to $Q^2\sim 40-50\;{\rm GeV}^2$~\cite{Schmookler:2022gxw}.
Over the past two decades, the high luminosity $e^+e^-$ colliders
has also started to explore the nucleon EMFFs in the time-like region~\cite{Antonelli:1998fv,BES:2005lpy,CLEO:2005tiu,Seth:2012nn,BaBar:2013ves,BaBar:2013ukx,BESIII:2021tbq}.

On the theoretical side, enormous intellectual endeavors have also been invested to unveiling the QCD dynamics beneath the nucleon EMFFs.
At small and moderate $Q^2$, the measured nucleon EMFFs have been confronted with the theoretical predictions made from various nonperturbative approaches,
{\it e.g.},
vector meson dominance model~\cite{Iachello:1972nu,Gari:1984ia,Lomon:2006xb,Brodsky:2003gs,Dubnickova:1992ii,Chen:2023oqs},
constitute quark model~\cite{Chung:1991st,Faessler:2005gd,deMelo:2008rj},  cloudy bag model~\cite{Thomas:1981vc,Betz:1983dy,Song:1992ei},
Dyson-Schwinger equation~\cite{Bashir:2012fs,Cloet:2013gva,Segovia:2014aza}, light-front Hamiltonian~\cite{Mondal:2019jdg},
chiral perturbation theory~\cite{Schindler:2005ke,Bauer:2012pv},
light-cone sum rule~\cite{Passek-Kumericki:2008uqr,Braun:2001tj}, lattice QCD~\cite{Collins:2011mk,Alexandrou:2011db,QCDSF:2017ssq,Capitani:2015sba,Alexandrou:2018sjm,Jang:2019jkn},
as well as dispersive analysis~\cite{Mergell:1995bf,Belushkin:2006qa}.

Thanks to asymptotic freedom,  the nucleon EMFFs with large momentum transfer are generally believed to
be adequately accounted by perturbative QCD (pQCD).
Within the collinear factorization formalism~\cite{Lepage:1979zb,Lepage:1979za,Lepage:1980fj,Efremov:1978rn,Efremov:1979qk,Duncan:1979ny,Duncan:1979hi,Chernyak:1977fk},
they can be expressed as the perturbatively-calculable hard-scattering kernel convoluted with nonperturbative
light-cone distribution amplitudes (LCDAs) of nucleon.
Helicity selection rule implies that, at asymptotically large $Q^2$,
$F_1$ scales with $1/Q^4$, and $F_2$ scales with $1/Q^6$~\cite{Brodsky:1974vy},
modulo logarithms of $Q^2$~\cite{Belitsky:2002kj}.
Hence nucleon Dirac form factor dominates Pauli form factor in such a limit.

The lowest-order (LO) hard-scattering kernel for $F_1$ was known in the late 70s,
shortly after the advent of QCD~\cite{Lepage:1979za,Chernyak:1980dj,Lepage:1980fj,Chernyak:1977fk}. Curiously, notwithstanding nearly half a centaury has elapsed,
the next-to-leading order (NLO) perturbative correction to this fundamental observable of nucleon remains elusive until today,
obviously due to severe technical obstacles~\footnote{Note the status of pQCD predictions for nucleons EMFFs lags far behind that for pion EMFFs,
where the one-loop QCD corrections were known in early 80s~\cite{Field:1981wx,Dittes:1981aw,Sarmadi:1982yg,Braaten:1987yy},
and even the two-loop corrections have also become available very recently~\cite{Chen:2023byr}.}.
The theoretical stagnation for such a basic observable over a very long time
is rather unusual, which should be contrasted with the $ep$ deep inelastic scattering,
where the three-loop QCD corrections became available about two decades ago~\cite{Vermaseren:2005qc}.

It is the very goal of this Letter to fill this overdue gap,
by accomplishing the calculation of the NLO perturbative correction
to the nucleon Dirac form factors, and conducting a comprehensive phenomenological investigation.

\begin{figure}[ht]
\centering
\includegraphics[width=0.5\textwidth]{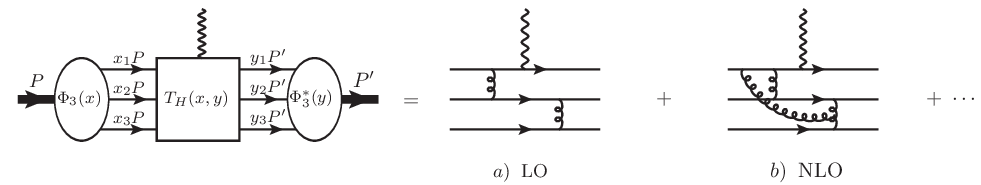}
\caption{Pictorial illustration of the nucleon EMFF in collinear factorization.
For concreteness, we show one typical LO and NLO Feynman diagrams for partonic reactions.}
\label{Feynman-diagrams}
\end{figure}

\vspace{0.2 cm}
\noindent{\color{blue}\it Collinear factorization for nucleon Dirac form factor.}  At large momentum transfer, the $F_1$
can be factorized as the following convolution integrals~\footnote{It is worth mentioning that,
this factorization formula is anticipated to be subject to modification starting at the two-loop order,
due to emergence of the new leading region contribution~\cite{Duncan:1979hi,Kivel:2010ns}.}
\bqa
Q^4 F_{1}(Q^2) = \Phi_3^*(y) \mathop{\otimes}_x T_H \left(x,y,Q^2\right) \mathop{\otimes}_y \Phi_3(x) +{\cal O}\left(\frac{1}{Q^2}\right),
\label{EMFF:collinear:factorization}
\eqa
where $\mathop{\otimes}\limits_{x}$ is a shorthand for the convolution integration
over $[dx]\equiv dx_1 dx_2 dx_3 \delta(1-\sum_i x_i)$. $T_H(x,y)$ signifies the hard-scattering kernel and $\Phi_3(x)$ represents the twist-3 nucleon LCDA.
This factorized structure is picturised in a cartoon in Fig.~\ref{Feynman-diagrams}.

For definiteness let us specialize to the proton. The leading-twist-3 LCDA of the proton assumes the following operator definition~\cite{Chernyak:1983ej,Braun:2000kw}:
\bqa
\langle 0\vert \epsilon_{ijk} ( u_{+}^{i\,T}(a_1 n) {\cal C} \not\! n u^j_{-}(a_2 n) ) \not \! n  d^k_{+}(a_3 n)\vert p^\uparrow (P)\rangle
\nn\\
\quad=-\frac{1}{2} P\cdot n \not\! n  u^\uparrow_{p}(P) \int[d x] e^{-i P\cdot n \sum x_i a_i} \Phi_3 (x),
\label{proton:LCDA:operator:def}
\eqa
where $i,j,k$ denote the color indices, and $q_\pm \equiv {1\over 2}(1\pm \gamma_5) q$
represents the right and left-handed components of the massless quark field,
and $u^\uparrow_{p}(P)$ signifies the Dirac spinor of $h={1\over 2}$ proton in relativistic normalization.
${\cal C}$ represents the charge conjugation matrix and
$n^\mu$ is a light-like vector. For simplicity we have suppressed those light-like gauge links whose role is to ensure gauge invariance.
Renormalizing the nonlocal baryonic operator in \eqref{proton:LCDA:operator:def},
one recovers the celebrated Efremov-Radyushkin-Brodsky-Lepage (ERBL) evolution equation~\cite{Lepage:1980fj, Efremov:1979qk}:
$
{d\Phi_3(x,\mu_F)}/{d\ln\mu^2_F} = V(x,y) \mathop{\otimes}\limits_{y} \Phi_3(y,\mu_F).
$

\vspace{0.2 cm}
\noindent{\color{blue}\it Perturbative matching in KM scheme.}
We first sketch how to achieve the desired factorization formula \eqref{EMFF:collinear:factorization}.
With $Q^2\gg \Lambda^2_{\rm QCD}$, asymptotic freedom indicates that
the transition matrix element in  \eqref{nucleon:formfactor} in $d$-dimensional spacetime
can be factorized as
\begin{eqnarray}
\langle \Pi^\mu (Q^2) \rangle & \equiv & \langle P' \vert j_{\rm em}^\mu
\vert P \rangle = \langle P' \vert {\bm[}\overline{\cal Q}_{\alpha\beta\gamma}(y){\bm]}^{\overline{\rm MS}} \vert 0 \rangle
\label{Pi_factorization:d:dimension}
\\
&& \;\mathop{\otimes}_y\;  {\cal T}_{\alpha\beta\gamma;\alpha'\beta'\gamma'}(x,y) \;\mathop{\otimes}_x\;
\langle 0\vert {\bm[}{\cal Q}_{\alpha'\beta'\gamma'}(x){\bm]}^{\overline{\rm MS}} \vert P\rangle,
\nn
\end{eqnarray}
where $\alpha, \beta, \gamma,\cdots$ denote the spinor indices affiliated with each quark line,
${\cal T}$ signifies the perturbative coefficient function. The three-quark
operator ${\cal Q}_{\alpha'\beta'\gamma'}(x)$ is defined via the Fourier transform of the non-local three-quark operators with light-like separations $\epsilon^{ijk} u^i_\alpha(a_1 n) u^j_\beta(a_2 n) d^k_\gamma(a_3 n)$, where the light-like gauge links tacitly inserted~\cite{Anikin:2013aka}.

It is advantageous to adopt the renormalization scheme originally proposed by Krankl and Manashov~\cite{Krankl:2011gch}
(KM scheme henceforth)
\footnote{The KM scheme has several remarkable virtues. It warrants the vanishing
of evanescent operators~\cite{Dugan:1990df} in $d=4$ dimensions so that one only needs to work on
the four-dimensional operator basis. The renormalization procedure in KM scheme preserves Fierz identities between the renormalized operators~\cite{Anikin:2013aka}.
The self-consistency of the KM scheme has been checked through the three-loop accuracy~\cite{Gracey:2012gx}.
It is also worth stressing that, the KM scheme has also been routinely utilized by the lattice QCD
practitioners to calculate the baryon LCDAs~\cite{Bali:2015ykx,RQCD:2019hps,Bali:2024oxg}.}.
The ${\overline{\rm MS}}$-renormalized three-quark operators in \eqref{Pi_factorization:d:dimension} is connected with
the bare one through
\beq
    {\bm[}{\cal Q}_{\alpha\beta\gamma}(x){\bm]}^{\overline{\rm MS}} = {\cal Z}_{\alpha\beta\gamma}^{\alpha'\beta'\gamma'}(x,x')
    \mathop{\otimes}_{x'} {\cal Q}_{\alpha'\beta'\gamma'}(x').
\label{Renormalization:three:quark:operator}
\eeq
For reader's convenience, the one-loop expression of ${\cal Z}$ in KM scheme is explicitly
given in the supplementary materials~\cite{Supplemental:Material}.

The perturbative coefficient function $\cal T$ in \eqref{Pi_factorization:d:dimension} can be decomposed into
\beq
    {\cal T}_{\alpha\beta\gamma;\alpha'\beta'\gamma'}(x,y) = T_{lmn}(x,y) \gamma^{(l)}_{\alpha\alpha'} \otimes \gamma^{(m)}_{\beta\beta'} \otimes \gamma^{(n)}_{\gamma\gamma'}.
\eeq
Note that $T_{lmn}$ is a scalar function, and Lorentz index $\mu$ is affiliated with one of the
$\gamma^{(l)}$, $\gamma^{(m)}$ and $\gamma^{(n)}$ matrices with $\gamma^{(n)}_{\mu_1\cdots\mu_n}
\equiv \gamma^{[\mu_1}\gamma^{\mu_2}\cdots\gamma^{\mu_{n}]}$~\cite{Krankl:2011gch,Anikin:2013aka}.

The central task now is to deduce the coefficient function $T_{lmn}(x,y)$.
Since it is insensitive to the long-distance physics, one can replace the incoming and
outgoing nucleons by ``fictitious" nucleons, {\it viz.},
the free, collinear three-quark states $|q_1(u_1P)q_2(u_2P)q_3(u_3P)\rangle$ and $|q_1(v_1P')q_2(v_2P')q_3(v_3P')\rangle$.
A virtue of the KM scheme is that the quark flavor structure
becomes irrelevant in determining ${\cal Z}$ and $T$~\cite{Krankl:2011gch}.

For a fictitious nucleon, the left-side in \eqref{Pi_factorization:d:dimension} can be computed in perturbation theory:
\beq
\langle \Pi^\mu(Q^2) \rangle = {\cal A}_{lmn}(u,v) \, \bar{u}_{\alpha\beta\gamma} \left[\gamma^{(l)}_{\alpha\alpha'} \otimes \gamma^{(m)}_{\beta\beta'} \otimes \gamma^{(n)}_{\gamma\gamma'} \right] u_{\alpha'\beta'\gamma'},
\label{Pi_current}
\eeq
where $u_{\alpha'\beta'\gamma'}\equiv u_{\alpha'}(u_1P) u_{\beta'}(u_2P) u_{\gamma'}(u_3P)$.

Analogously, we also evaluate the nonlocal matrix elements in the right-side of \eqref{Pi_factorization:d:dimension} perturbatively.
After dropping all the scaleless integrals, we obtain
\beq
    \langle 0\vert {\bm[}{\cal Q}_{\alpha\beta\gamma}(x){\bm]}^{\overline{\rm MS}} \vert P\rangle
= {\cal Z}_{\alpha\beta\gamma}^{\alpha'\beta'\gamma'}(x,u) \,  u_{\alpha'\beta'\gamma'}.
\eeq

Practically we utilize the following matching equation to deduce $T$~\footnote{
Note that one has to reexpand the products of $\gamma$-marices in the right-hand side  of \eqref{Practical:one:loop:matching:formula},
and finally cast them in the basis of the totally anti-symmetrized products of $\gamma$-matrices.
}:
\bqa
 && {\cal A}_{lmn}(u,v) \gamma^{(l)}_{\alpha\alpha'} \otimes \gamma^{(m)}_{\beta\beta'} \otimes \gamma^{(n)}_{\gamma\gamma'}  = { \overline {\cal Z} }_{\alpha\beta\gamma}^{\alpha_2\beta_2\gamma_2}(y,v)
 \label{Practical:one:loop:matching:formula}
 \\
 && \mathop{\otimes}_{y} \left[ T_{l'm'n'}(x,y) \gamma^{(l')}_{\alpha_2\alpha_1} \otimes \gamma^{(m')}_{\beta_2\beta_1} \otimes \gamma^{(n')}_{\gamma_2\gamma_1} \right] \mathop{\otimes}_{x} {\cal Z}_{\alpha'\beta'\gamma'}^{\alpha_1\beta_1\gamma_1}(x,u). \nn
\eqa

\vspace{0.2 cm}

\noindent{\color{blue}\it Sketch of the calculation.}
We start with the partonic reaction $\gamma^*+q_1 q_2 q_3 \to
q_1 q_2 q_3$.
We use the packages {\tt HepLib}~\cite{Feng:2021kha} and {\tt FeynArts}~\cite{Hahn:2000kx} to generate the Feynman diagrams and
the corresponding amplitudes. There are about 48 tree-level diagrams, and 2040 one-loop diagrams.

As is evident in Fig.~\ref{Feynman-diagrams}, the amplitude in each diagram bears the form
$\bar{u}_{\alpha\beta\gamma} \left[ \Gamma^{(1)}_{\alpha\alpha'} \otimes \Gamma^{(2)}_{\beta\beta'} \otimes \Gamma^{(3)}_{\gamma\gamma'} \right] u_{\alpha'\beta'\gamma'}$.
Moving $P\!\!\!\!/$ to the rightmost and $P\!\!\!\!/\,'$ to the leftmost  repeatedly using the anti-communication relation~\footnote{Note that
$\ell\!\!\!/$ also appears ubiquitously in the intermediate stage. By Lorentz covariance, one can trade the loop momentum $\ell$ for
the metric tensor or the external momenta $P$ and $P'$.},
and employing Dirac equation, each $\Gamma^{(i)}$ can be expressed as the product of a string of Dirac $\gamma$ matrices,
which can be further cast into the form prescribed in \eqref{Pi_current}.

The aforementioned operations can be automated by  {\tt HepLib}~\cite{Feng:2021kha}.
We then use {\tt Apart}~\cite{Feng:2012iq} and {\tt FIRE}~\cite{Smirnov:2019qkx} for partial fraction and IBP reduction.
We end up with about 100 one-loop two-point and three-point master integrals, which are computed using
the differential equation method~\cite{Kotikov:1990kg,Remiddi:1997ny}.
Upon renormalizing the QCD coupling in $\overline{\rm MS}$ scheme, we obtain the UV-finite expression of the NLO coefficients
${\cal A}_{lmn}$ in \eqref{Pi_current}, which however still possess uncancelled single infrared pole.
Validated by the matching equation \eqref{Practical:one:loop:matching:formula},
we find that, hearteningly, the IR poles in ${\cal A}_{lmn}$ are exactly cancelled by the corresponding poles in ${\cal Z}$,
so that the short-distance coefficients $T_{lmn}(x,y)$ are indeed IR finite in a pointwise manner.
Therefore we verify that the collinear factorization does hold at one-loop order
for the nucleon Dirac form factor.

To recover the desired structure in \eqref{EMFF:collinear:factorization},
we substitute the following expression~\cite{Braun:1999te,Anikin:2013aka}
\bqa
&&\hspace{-0.3cm}\langle 0\vert {\bm[}{\cal Q}_{\alpha\beta\gamma}(x){\bm]}^{\overline{\rm MS}} \vert P\rangle
= \frac{1}{4} \Big\{ {\cal V}(P\!\!\!\!/ {\cal C})_{\alpha\beta} (\gamma_5 N)_\gamma
\\
&&\hspace{1cm} + {\cal A}(P\!\!\!\!/\gamma_5 {\cal C})_{\alpha\beta}N_\gamma +{\cal T}(i\sigma_{\mu P}{\cal C})_{\alpha\beta} (\gamma_\mu\gamma_5 N)_\gamma \Big\}
\nn
\eqa
into \eqref{Pi_factorization:d:dimension}. Using the relations among
${\cal V}(x)$, ${\cal A}(x)$, ${\cal T}(x)$ and $\Phi_3(x)$~\cite{Braun:1999te,Anikin:2013aka},
we obtain $Q^4 \langle \Pi^\mu(Q^2) \rangle = \Phi_3(x) \mathop{\otimes}\limits_x T_H(x,y)\mathop{\otimes}\limits_y \Phi_3^*(y) \, \bar{N} \gamma^\mu N$.
The intended Dirac form factor through NLO accuracy can then be predicted in line  with \eqref{EMFF:collinear:factorization}.

\vspace{0.2 cm}

\noindent{\color{blue}\it Master formula for nucleon Dirac form factor at NLO.}
The leading-twist nucleon LCDA is conveniently expanded in the orthogonal polynomials basis~\cite{Braun:2008ia}:
\beq
\Phi_3(x,\mu_F) =  f_{\rm N}(\mu_F) \!\!\sum_{0\le i\le j}\!\!  a_{ij}(\mu_F) \phi_{ij}(x),
\label{neuclon:DA:expansion}
\eeq
with $\phi_{ij}(x) = 120 x_1 x_2 x_3 \mathcal{P}_{ij}(x)$ and $f_{\rm N}$ denotes the nucleon decay constant. The first few orthogonal polynomials ${\cal P}_{ij}(x)$ have been tabulated in \cite{Braun:2008ia,Braun:2014wpa}({\it e.g}, (12) in \cite{Braun:2014wpa}).
All the nonperturbative dynamics is encoded in the expansion coefficients $a_{ij}(\mu_F)$.

For phenomenological usage, it is advantageous to recast \eqref{EMFF:collinear:factorization}
from the convolutional form into an algebraic one:
\bqa
Q^4 F_1(Q^2) = f_{\rm N}^2(\mu_F) {\sum_{ijkl}} a_{ij}(\mu_F) {\cal H}^{ijkl} a_{kl}(\mu_F),
\label{Master:formula:nucleon:EM:FF}
\eqa
with ${\cal H}^{ijkl} \equiv \phi_{ij}(x) \mathop{\otimes}\limits_{x} T_H(x,y) \mathop{\otimes}\limits_{y} \phi_{kl}(y)$.

The short-distance coefficients ${\cal H}^{ijkl}$ can be organized
in powers of $\alpha_s$.
Through NLO accuracy, they can be parameterized as
\bqa
{\cal H}^{ijkl} &=& \frac{1600 \pi^2 \alpha_s^2}{3} \Big\{ e_u\, c^{ijkl}_0 \left[ 1+ {\alpha_s\over \pi}\left( c^{ijkl}_1 L_\mu + c^{ijkl}_2 \right) \right]
\nn\\
&&\hspace{-0.7cm} +  e_d\,d^{ijkl}_0 \left[1+ {\alpha_s\over \pi} (d^{ijkl}_1 L_\mu+d^{ijkl}_2) \right] + {\cal O}(\alpha_s^2) \Big\},
\label{short:dist:Tmnm'n':def}
\eqa
with $L_\mu \equiv \ln \mu^2/Q^2$. For simplicity we have set $\mu_R=\mu_F=\mu$. Note
${\cal H}^{ijkl}={\cal H}^{klij}$ and $c^{ijkl}_1=d^{ijkl}_1$.
For a given $ijkl$, the coefficients  $c^{ijkl}_{0,1,2}$ and $d^{ijkl}_{0,1,2}$ can be analytically inferred, whose
explicit expressions are given in the supplementary materials~\cite{Supplemental:Material}.

\begin{table}[h]
\centering\setlength{\tabcolsep}{2mm}{}
\begin{tabular}{c|cc|cc|c}
\hline\hline
${ijkl}$ & $c^{ijkl}_0$ & $c^{ijkl}_2$ & $d^{ijkl}_0$ & $d^{ijkl}_2$ & $c^{ijkl}_1=d^{ijkl}_1$ \\
\hline\hline
$0000$ & 1.0000 & 21.334 & 2.0000 & 17.491 & 4.8333 \\
\hline
$1000$ & 8.1667 & 24.347 & -8.1667 & 24.347 & 5.3889 \\
$1010$ & 168.78 & 28.623 & 174.22 & 28.251 & 5.9444 \\
\hline
$1100$ & 3.5000 & 28.699 & -3.5000 & 18.233 & 5.5000  \\
$1110$ & 38.111 & 29.655 & -38.111 & 29.655 & 6.0556 \\
$1111$ & 59.889 & 27.016 & 21.778 & 26.056 & 6.1667 \\
\hline
$2000$ & 21.350 & 26.260 & 8.0500 & 24.454 & 5.7222 \\
$2010$ & 89.833 & 31.367 & -89.833 & 31.367 & 6.2778 \\
$2011$ & 93.100 & 32.158 & 9.8000 & 37.615 & 6.3889 \\
$2020$ & 234.71 & 34.390 & 54.880 & 33.494 & 6.6111 \\
\hline
$2100$ & 8.7500 & 27.639 & -8.7500 & 27.639 & 5.9444 \\
$2110$ & 159.25 & 32.756 & 159.25 & 32.904 & 6.5000 \\
$2111$ & 44.917 & 32.687 & -44.917 & 32.687 & 6.6111 \\
$2120$ & 84.525 & 35.895 & -84.525 & 35.895 & 6.8333 \\
$2121$ & 159.25 & 37.626 & 159.25 & 37.603 & 7.0556  \\
\hline
$2200$ & 3.2000 & 27.381 & 0.10000 & 45.283 & 6.0000 \\
$2210$ & 12.250 & 31.624 & -12.250 & 31.624 & 6.5556 \\
$2211$ & 7.1167 & 39.349 & 2.6833 & 35.342 & 6.6667 \\
$2220$ & 30.345 & 35.977 & 4.9350 & 36.841 & 6.8889 \\
$2221$ & 7.1750 & 38.780 & -7.1750 & 38.780 & 7.1111 \\
$2222$ & 7.7900 & 34.975 & 1.7200 & 34.974 & 7.1667 \\
\hline\hline
\end{tabular}
\caption{Numerical values of $c_i$ and $d_i$ affiliated with various ${\cal H}^{ijkl}$ with $0\le \{ijkl\} \le 2$.}
\label{tab:Tmn-numeric-values}
\end{table}

We conduct the four-fold convolution integration following two different routes:
one is to utilize the numerical integrator {\tt Cubature}~\cite{cubature}
by implementing multiple-precision floating-point around 100 digits, and the other is to
carry out the integration analytically with the assistance of the package {\tt PolyLogTools}~\cite{Duhr:2019tlz}.
The perfect agreement is found between the analytic and numerical approaches.
In Table~\ref{tab:Tmn-numeric-values} we enumerate the values for $c_i$ and $d_i$ associated with
some low-lying $(ijkl)$~\footnote{We notice that, a PhD thesis by Kn\"{o}dlseder in 2015 was
dedicated to calculating the NLO QCD correction to nucleon Dirac form factor at leading twist~\cite{Knodlseder:2015vmu}.
Though the analytic result of the one-loop amplitude was obtained in \cite{Knodlseder:2015vmu},
the IR-finite hard-scattering kernel $T^{(1)}(x,y)$ has not been given in a pointwise manner.
More importantly, since the author failed to accomplish the four-fold convolution integrations
in a numerically stable fashion, no attempt of confronting the NLO pQCD predictions with the measured nucleon form factor
data was made in \cite{Knodlseder:2015vmu}.}.

Equipped with Table~\ref{tab:Tmn-numeric-values},
Equation~\eqref{Master:formula:nucleon:EM:FF} constitutes the master formula of rendering
predictions for proton Dirac form factor through one-loop accuracy.
In the isospin symmetric limit, the neutron Dirac form factor $F_1^n$
can be obtained from \eqref{Master:formula:nucleon:EM:FF} by making the replacement
$e_u \leftrightarrow e_d$ in \eqref{short:dist:Tmnm'n':def}.
It is also straightforward to adapt this master formula from the space-like domain to the time-like one,
provided that one makes the replacement $L_\mu \to  L_\mu+ i\pi$ in \eqref{short:dist:Tmnm'n':def},
with $Q^2$ now indicating the squared invariant mass of the nucleon-antinucleon pair.

\vspace{0.2 cm}

\noindent{\color{blue}\it Phenomenological exploration.}
It is now the time to confront the state-of-the-art pQCD predictions with the available large-$Q^2$  data.
We take the data points for spacelike proton Dirac form factor from \cite{Coward:1967au,Sill:1992qw}~\footnote{Note it is the magnetic Sachs form factor of nucleon, $G_{\rm M}(Q^2)$,
that is directly accessible in $ep$ scattering experiments.
At large $Q^2$, it nevertheless becomes legitimate to assume $F_1(Q^2)\approx G_{\rm M}(Q^2)$.},
and the data points for time-like proton Dirac form factor from \cite{BES:2005lpy,CLEO:2005tiu,Seth:2012nn,E835:1999mlt,E760:1992rvj,BaBar:2013ves,BaBar:2013ukx}.
We discard many irrelevant data at small momentum transfer.
In contrast, the quality of data and the covered range of $Q^2$ for neutron EMFFs
is much less competitive with respect to that for proton.
The data points for spacelike neutron Dirac form factor are taken from \cite{Rock:1982gf,Lung:1992bu},
while those for timelike neutron Dirac form factor are taken from \cite{Antonelli:1998fv,BESIII:2021tbq}.

\begin{figure}[htb]
\centering
\includegraphics[width=0.5\textwidth]{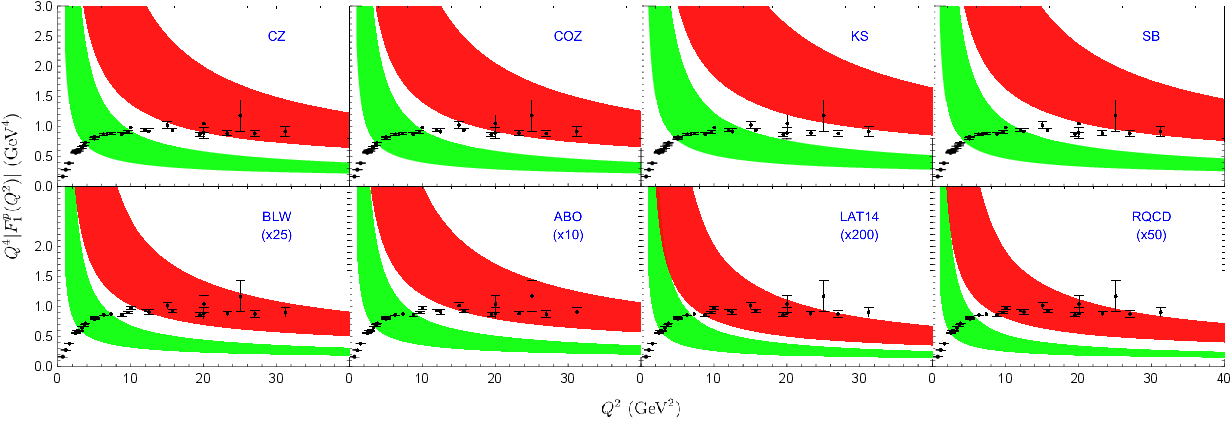}
\caption{pQCD predictions v.s. data for $Q^4 F^p_1(Q^2)$ in the space-like region.
The values of the expansion coefficients of proton LCDA are taken with eight different ansatzs.
The green and red bands correspond to the LO and NLO pQCD predictions,
which are obtained by sliding $\mu_R=\mu_F=\mu$ from $Q/2$ to $Q$.
Since the pQCD predictions in the lower row are much smaller than the data, we have multiplied them with
a magnifying factor (as specified inside the parenthesis in each plot) to render the theoretical bands visible.}
\label{Plot:spacelike:proton:EMFF}
\end{figure}

\begin{figure}[htb]
\centering
\includegraphics[width=0.5\textwidth]{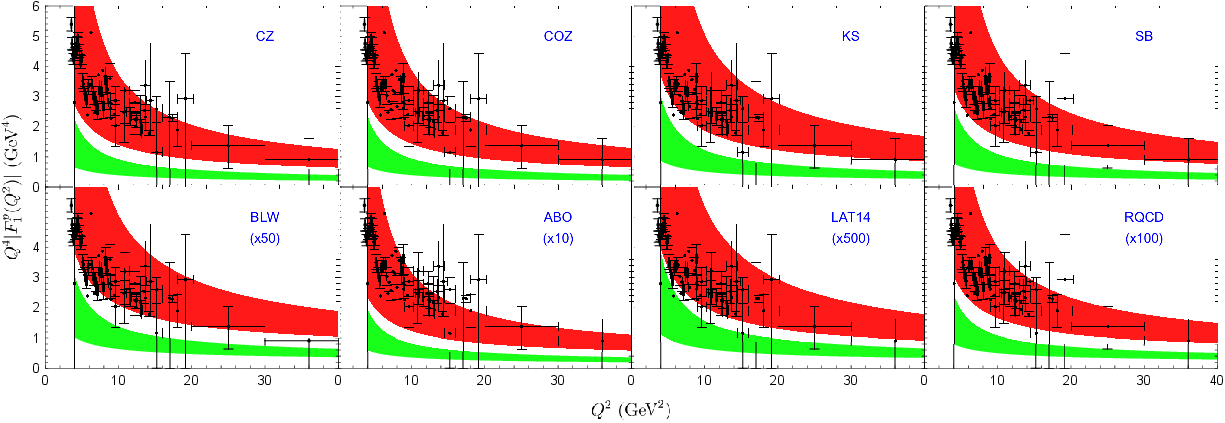}
\caption{Same as Fig.~\ref{Plot:spacelike:proton:EMFF}, except switching from space-like to time-like region. }
\label{Plot:timelike:proton:EMFF}
\end{figure}

\begin{figure}[htb]
\centering
\includegraphics[width=0.5\textwidth]{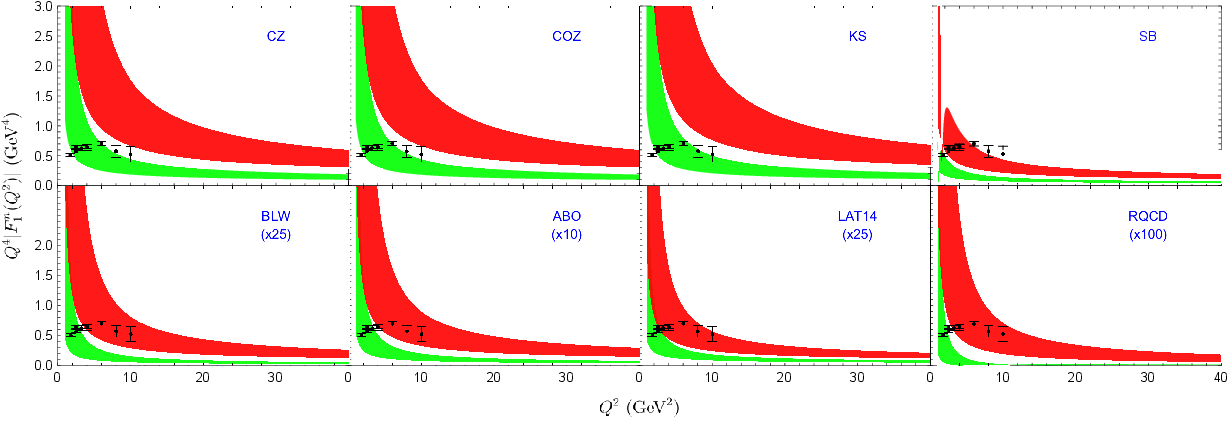}
\caption{Similar to Fig.~\ref{Plot:spacelike:proton:EMFF}, but with proton replaced with the neutron.
Here pQCD predictions are confronted with the data for $Q^4 F^n_1(Q^2)$ in the space-like region. }
\label{Plot:spacelike:neutron:EMFF}
\end{figure}

\begin{figure}[htb]
\centering
\includegraphics[width=0.5\textwidth]{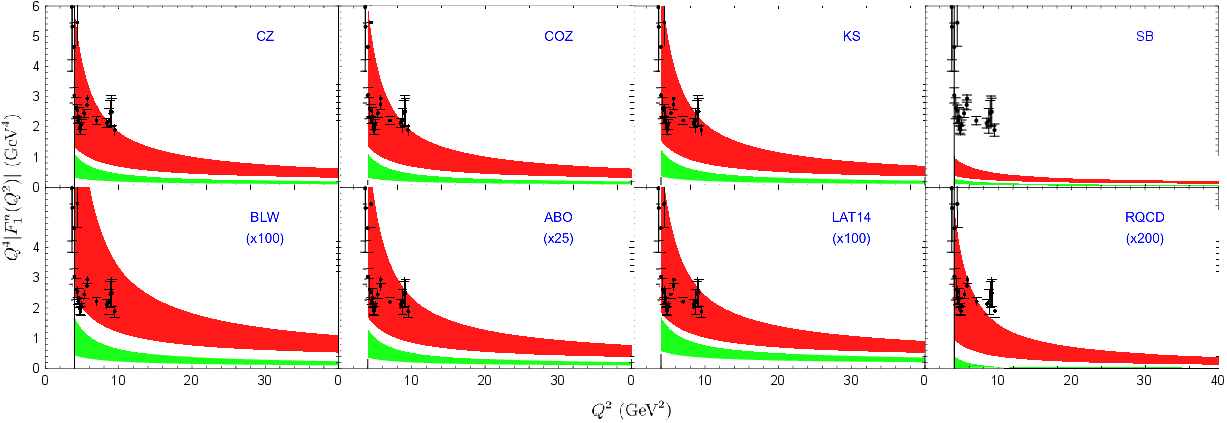}
\caption{Same as Fig.~\ref{Plot:spacelike:neutron:EMFF}, except switching from space-like to time-like region.}
\label{Plot:timelike:neutron:EMFF}
\end{figure}

As the key nonperturative input, our present knowledge about the nucleon LCDAs is far from confirmative.
During the past decades, the low-lying expansion coefficients of nucleon LCDAs have been extensively investigated by different approaches~\cite{Chernyak:1984bm,Chernyak:1987nv,King:1986wi,Stefanis:1992nw,Braun:2006hz,Anikin:2013aka,Braun:2014wpa,RQCD:2019hps}.
In our numerical analysis, we have taken eight different ansatzs of nucleon LCDAs as input.
Six of them are obtained from the QCD sum rules, referred to as {\tt CZ}~\cite{Chernyak:1984bm}, {\tt COZ}~\cite{Chernyak:1987nu}, {\tt KS}~\cite{King:1986wi}, {\tt SB}~\cite{Stefanis:1992nw}, {\tt BLW}~\cite{Braun:2006hz} and {\tt ABO}~\cite{Anikin:2013aka}, respectively.
Two of them are provided by the recent lattice QCD simulations,
abbreviated as {\tt LAT 14}~\cite{Braun:2014wpa} and {\tt RQCD}~\cite{RQCD:2019hps}, respectively.

The predicted values of the expansion coefficients of nucleon are usually given at the initial scale $\mu_0=1$ or 2 GeV,
which are scattered in a wide range in literature.
The one-loop renormalization group equations are invoked to evolve $f_{\rm N}$ and $a_{ij}(\mu)$ from $\mu_0$ to another scale,
{\it viz.}, $f_{\rm N}(\mu)=f_{\rm N}(\mu_0)\left[\alpha_s(\mu)/\alpha_s(\mu_0)\right]^{2/(3\beta_0)}$ and
$a_{ij}(\mu)=a_{ij}(\mu_0) \left[\alpha_s(\mu)/\alpha_s(\mu_0)\right]^{\gamma_{ij}/\beta_0}$.
$\beta_0=11-{2\over 3}n_f$ with $n_f=3$ signifies the one-loop coefficient of the QCD $\beta$ function,
while $\gamma_{ij}$ is the anomalous dimension associated with each expansion coefficient,
whose explicit form can be found in \cite{Braun:2014wpa}.
The package {\tt FAPT}~\cite{Bakulev:2012sm} is utilized to evaluate the running QCD coupling constant
to two-loop accuracy.

From Fig.~\ref{Plot:spacelike:proton:EMFF} through Fig.~\ref{Plot:timelike:neutron:EMFF}, we confront our state-of-the-art pQCD predictions
with the available large-$Q^2$ data for proton and neutron's Dirac form factors, in both space-like and time-like cases.
The theoretical errors are estimated by varying $\mu_R=\mu_F=\mu$ from $Q/2$ to $Q$.

It has been widely known that the LO predictions in pQCD for proton Dirac form factor are generally
much smaller than the large-$Q^2$ data, for virtually all the popular models of nucleon LCDAs~\cite{Stefanis:1997zyh}
\footnote{To resolve this discrepancy, historically it has been proposed~\cite{Chernyak:1984bm,Gari:1986ue,Stefanis:1987vr,Gari:1986dr,Ji:1986uh,Stefanis:1992nw}
that the strong coupling constants should
be evaluated at the exact gluon virtualities in each diagram, so that the LO prediction can be significantly
enhanced to account for the data. The $\alpha_s$ in this recipe might be associated with a scale of order $\Lambda_{\rm QCD}$,
evaluation of which requires some ad hoc treatment. Moreover, it is difficult to adapt this recipe to the NLO case.}.
As is evident from Fig.~\ref{Plot:spacelike:proton:EMFF} through Fig.~\ref{Plot:timelike:neutron:EMFF},
the magnitudes of the NLO QCD correction are positive and significant.
Once we adopt the {\tt CZ}~\cite{Chernyak:1984bm}, {\tt COZ}~\cite{Chernyak:1987nu}, {\tt KS}~\cite{King:1986wi}, {\tt SB}~\cite{Stefanis:1992nw}
parametrizations of nucleon LCDAs, the pQCD predictions at NLO accuracy appear to be  in magnitudes compatible with the
large-$Q^2$ data for both proton and neutron's Dirac form factors, in both spacelike and timelike regions.
Nevertheless, we must stress that the monotonically decreasing tendency of the NLO predictions in variation with $Q^2$
is clearly at odds with that exhibited in the actual data.

In sharp contrast, due to quite small expansion coefficients predicted by {\tt BLW}~\cite{Braun:2006hz}, {\tt ABO}~\cite{Anikin:2013aka},
{\tt LAT14}~\cite{Braun:2014wpa} and {\tt RQCD}~\cite{RQCD:2019hps}, the resulting pQCD predictions
fall far below the experimental data.
From Fig.~\ref{Plot:spacelike:proton:EMFF} through Fig.~\ref{Plot:timelike:neutron:EMFF}, one clearly sees that,
even after including NLO corrections, the pQCD predictions based on these parametrizations
are still order-of-magnitude smaller than the measured Dirac form factor of proton and neutron.

Since lattice QCD is universally believed to be the only model-independent nonperturbative approach to
unravel the hadronic structure, the sheer failure of the parameterized nucleon LCDAs provided by {\tt LAT14} and {\tt RQCD}
when confronting with the data is rather disquieting.
To settle this alarming discrepancy, it seems imperative to call for another independent lattice study
based on different algorithm. The large momentum effective theory (LaMET)~\cite{Ji:2013dva,Ji:2020ect} is a viable option,
which allows one to ascertain hadron's parton distributions directly in the $x$ space on Euclidean lattice
Some theoretical exploration to infer light octet baryons' LCDAs  in LaMET approach has recently been launched~\cite{Deng:2023csv,Han:2024ucv}.
Had the prospective LaMET predictions for nucleon LCDAs turned out to be compatible with the {\tt RQCD} predictions,
one would be then inevitably led to the conclusion that the hard-scattering mechanism alone in \eqref{EMFF:collinear:factorization} would be
inadequate even at $Q^2\approx 30\;{\rm GeV^2}$,  and some additional
soft nonfactorizable contributions are expected to play an important role.

\vspace{0.2 cm}

\noindent{\color{blue}\it Summary and Outlook.}
In this work we report the first complete calculation of the NLO QCD corrections to the nucleon Dirac form factors.
The collinear factorization formalism has been explicitly verified to hold true
at one-loop order for hard-exclusive process involving baryons.

It has long been known that the LO predictions in pQCD severely underestimate the measured nucleon Dirac form factor data.
Historically some ad hoc frozen strong coupling recipe were advocated to reconcile this discrepancy.
Remarkably, the effect of NLO perturbative corrections turn to be positive and significant. 
Nevertheless, even after including the sizable NLO QCD corrections, the pQCD predictions based on the recent lattice QCD inputs of
nucleon LCDAs, are still considerably smaller than the measured nucleon form factors. 
This symptom may indicate that the hard-scattering mechanism alone might be inadequate even at $Q^2$ accessible in future {\tt EIC/EicC} experiments,  
and other soft nonfactorizable contributions need be supplemented to account for the data.

In a sense, this work may herald the coming of a new phase of perturbative QCD,
where NLO QCD corrections start to be systematically explored for a class of important hard exclusive reactions in baryon sector,
exemplified by the nucleon Pauli (axial, gravitational) form factors,
nucleon Compton scattering, $\gamma\gamma\to N\overline{N}$, quarkonium decay into $N\overline{N}$, and so on.
It is conceivable that the effect of NLO QCD corrections could be significant in many occasions,
including which will thereby facilitate a critical comparison between pQCD prediction and data.

\vspace{0.2 cm}
\noindent{\color{blue}\it Acknowledgments.}
We are grateful to Deshan Yang for valuable discussion.
L.~B.~C. is in part supported by National Natural Science Foundation of China (NSFC) under Grant No.~12175048 and the Guangdong Basic and Applied Basic Research Foundation under grant No.~2025B1515020009.
W.~C. is supported by NSFC under Grant No.~12405095 and Guangdong Major Project of Basic and Applied Basic Research~(No. 2020B0301030008).
The work of F.~F. is supported by NSFC under Grant No.~12275353 and in part by the Fundamental Research Funds for the Central Universities~(2025 Basic Sciences Initiative in Mathematics and Physics).
The work of S.~H. and Y.~J. is supported in part by NSFC under Grants No.~12475090 and No.~11925506.

\vspace{0.2 cm}

\noindent{\color{blue}\it Note added.}
After the first version of this paper was submitted to arXiv, there has appeared a similar work by Huang, Shi, Wang and Zhao~\cite{Huang:2024ugd}, who have also computed the NLO QCD correction to nucleon Dirac form factor in the spacelike domain, yet within a specific evanescent-to-physical operator mixing scheme. After converted into KM scheme, their results in the supplementary material in the published version of \cite{Huang:2024ugd} are compatible with ours.


\appendix

\begin{widetext}

\section{Supplementary Material }

\subsection{One-loop renormalization factor of  non-local three-quark operators in KM scheme}

In KM scheme~\cite{Krankl:2011gch}, The ${\overline{\rm MS}}$-renormalized three-quark operators is connected with the bare one through
\beq
    {\bm[}{\cal Q}_{\alpha\beta\gamma}(x){\bm]}^{\overline{\rm MS}} = {\cal Z}_{\alpha\beta\gamma}^{\alpha'\beta'\gamma'}(x,x')
    \mathop{\otimes}_{x'} {\cal Q}_{\alpha'\beta'\gamma'}(x').
\label{Renormalization:three:quark:operator}
\eeq
where the ${\cal Z}$ factor can be parameterized as
\bqa
{\cal Z}_{\alpha\beta\gamma}^{\alpha'\beta'\gamma'}(x,x') &=& \delta_{\alpha\beta\gamma}^{\alpha'\beta'\gamma'}\delta(x-x')
+\sum_{s=1} {1\over \epsilon^{s} }\, a^{s}_{lmn} \, \gamma^{(l)}_{\alpha\alpha'} \otimes \gamma^{(m)}_{\beta\beta'} \otimes \gamma^{(n)}_{\gamma\gamma'}
\eqa
where $\gamma^{(n)}$ is the shorthand for the totally anti-symmetrized product of $n$ $\gamma$-matrices,
$\gamma^{(n)}_{\mu_1\cdots\mu_n}  \equiv \gamma^{[\mu_1}\gamma^{\mu_2}\cdots\gamma^{\mu_{n}]}$.
Note that all the Lorentz indices of $\gamma$-matrices are contracted between different quark lines,
where there is only one independent way to contract all indices in ${\cal Z}$.

To the best of our knowledge, the explicit one-loop expression for the renormalization factor ${\cal Z}$ in
Eq.(5),
which is associated with the nonlocal three-quark operators with light-like separations,
has never been explicitly given in literature. For reader's convenience, here we present its closed one-loop form:

\begin{eqnarray}
{\cal Z}&=& \delta(x_1-x'_1)\delta(x_2-x'_2)\Gamma_{000}-\frac{\alpha_s}{12 \pi} \frac{1}{\epsilon} \Bigg\{
18 \delta(x_1-x'_1) \delta(x_2-x'_2) \Gamma_{000}
\nonumber\\
&& \left. +\Gamma_{000}\left\{ \delta(x'_3-x_3)\left[\frac{4 x_2}{x'_2} \left[\frac{\theta(x_1-x'_1)}{x_1-x'_1}\right]_+ + \frac{4 x_1}{x'_1} \left[\frac{\theta(x_2-x'_2)}{x_2-x'_2}\right]_+-\frac{4(x'_1-x_1) \theta(x_2-x'_2)}{x'_1 x'_2}+\frac{4 x_2}{x'_2(1-x'_3)}\right]\right.\right.
\nonumber\\
&& +\delta(x'_2-x_2)\left[\frac{4 x_1}{x'_1(1-x'_2)}+\frac{4(x'_1-x_1) \theta(x_1-x'_1)}{x'_1 x'_3}+\frac{4 x_1}{x'_1} \left[\frac{\theta(x_3-x'_3)}{x_3-x'_3}\right]_+ + \frac{4 x_3}{x'_3} \left[\frac{\theta(x_1-x'_1)}{x_1-x'_1}\right]_+ \right]
\nonumber\\
&& \left.+\delta(x'_1-x_1)\left[\frac{4 x_2}{(1-x'_1) x'_2}+\frac{4(x'_2-x_2) \theta(x_2-x'_2)}{x'_2 x'_3}+
\frac{4 x_2}{x'_2} \left[\frac{\theta(x_3-x'_3)}{x_3-x'_3}\right]_+ + \frac{4 x_3}{x'_3} \left[\frac{\theta(x_2-x'_2)}{x_2-x'_2}\right]_+ \right]\right\}
\nonumber\\
&& +\delta(x'_2-x_2) \Gamma_{202}\left[\frac{(x'_1-x_1) \theta(x_1-x'_1)}{x'_1 x'_3}+\frac{x_1}{x'_1(1-x'_2)}\right]
+\delta(x'_1-x_1) \Gamma_{022}\left[\frac{(x'_2-x_2)\theta(x_2-x'_2)}{x'_2 x'_3}+\frac{x_2}{(1-x'_1) x'_2}\right]
\nonumber\\
&& +\delta(x'_3-x_3) \Gamma_{220}\left[\frac{x_2}{x'_2(1-x'_3)}-\frac{(x'_1-x_1) \theta(x_2-x'_2)}{x'_1 x'_2}\right]
\Bigg\}
+{\cal O}(\alpha_s^2),
\end{eqnarray}
where the basis matrices $\Gamma_{ijk}$ are defined by
\begin{eqnarray}
\Gamma_{000} = I\otimes I\otimes I, \quad
\Gamma_{220} = \gamma^{(2)}_{\mu\nu} \otimes \gamma^{(2)}_{\mu\nu} \otimes I, \quad
\Gamma_{202} = \gamma^{(2)}_{\mu\nu}  \otimes I \otimes \gamma^{(2)}_{\mu\nu}, \quad
\Gamma_{022} = I \otimes \gamma^{(2)}_{\mu\nu}  \otimes \gamma^{(2)}_{\mu\nu}.
\end{eqnarray}

The $+$-distribution is defined by
\begin{eqnarray}
    \int_0^1 dx_i \left[\frac{\theta(x_i-x'_i)}{x_i-x'_i}\right]_+ f(x_i) = \int_0^1 dx_i \frac{\theta(x_i-x'_i)}{x_i-x'_i} \left[ f(x_i) - f(x'_i) \right],
\end{eqnarray}
where $f(x_i)$ signifies a test function.

Moreover, with the aid of the following identities,
\begin{eqnarray}
&&\frac{x_2}{x'_2} \left[ \frac{\theta(x_1-x'_1)}{x_1-x'_1} \right]_+ = \left[ \frac{x_2}{x'_2} \frac{\theta(x'_2-x_2)}{x'_2-x_2} \right]_+ - \delta(x_1-x'_1),
\nn\\
&&\frac{x_2}{x'_2(1-x'_3)}-\frac{(x'_1-x_1)\theta(x_2-x'_2)}{x'_1 x'_2}
= \left[\frac{x_2}{x'_2}\theta(x'_2-x_2) + \frac{x_1}{x'_1}\theta(x'_1-x_1)\right] \frac{1}{(1-x'_3)},
\end{eqnarray}
one can obtain an alternative expression for ${\cal Z}$:
\begin{eqnarray}
&&\hspace{-0.5cm} {\cal Z} = \delta(x_1-x'_1)\delta(x_2-x'_2)\Gamma_{000}-\frac{\alpha_s}{12 \pi} \frac{1}{\epsilon} \Bigg\{
-6 \delta(x_1-x'_1) \delta(x_2-x'_2) \Gamma_{000}
\nonumber\\
&& +\Gamma_{000}\left\{ \delta(x'_3-x_3)\left\{ \left[\frac{4 x_2}{x'_2}\frac{\theta(x'_2-x_2)}{x'_2-x_2} + \frac{4 x_1}{x'_1}\frac{\theta(x'_1-x_1)}{x'_1-x_1}\right]_+
+\left[\frac{4 x_2}{x'_2}\theta(x'_2-x_2)+\frac{4 x_1}{x'_1}\theta(x'_1-x_1)\right]\frac{1}{1-x'_3}\right\}\right.
\nonumber\\
&& +\delta(x'_2-x_2)\left\{ \left[\frac{4 x_3}{x'_3}\frac{\theta(x'_3-x_3)}{x'_3-x_3} + \frac{4 x_1}{x'_1}\frac{\theta(x'_1-x_1)}{x'_1-x_1}\right]_+
+\left[\frac{4 x_3}{x'_3}\theta(x'_3-x_3)+\frac{4 x_1}{x'_1}\theta(x'_1-x_1)\right]\frac{1}{1-x'_2}\right\}
\nonumber\\
&& +\delta(x'_1-x_1)\left\{ \left[\frac{4 x_2}{x'_2}\frac{\theta(x'_2-x_2)}{x'_2-x_2} + \frac{4 x_3}{x'_3}\frac{\theta(x'_3-x_3)}{x'_3-x_3}\right]_+
+\left[\frac{4 x_2}{x'_2}\theta(x'_2-x_2)+\frac{4 x_3}{x'_3}\theta(x'_3-x_3)\right]\frac{1}{1-x'_1}\right\}
\nonumber\\
&& +\delta(x'_3-x_3) \left[\frac{x_2}{x'_2}\theta(x'_2-x_2)+\frac{x_1}{x'_1}\theta(x'_1-x_1)\right]\frac{\Gamma_{220}}{1-x'_3}
+\delta(x'_2-x_2) \left[\frac{x_3}{x'_3}\theta(x'_3-x_3)+\frac{x_1}{x'_1}\theta(x'_1-x_1)\right]\frac{\Gamma_{202}}{1-x'_2}
\nonumber\\
&&
+\delta(x'_1-x_1) \left[\frac{x_2}{x'_2}\theta(x'_2-x_2)+\frac{x_3}{x'_3}\theta(x'_3-x_3)\right]\frac{\Gamma_{022}}{1-x'_1}
\Bigg\}+{\cal O}(\alpha_s^2).
\end{eqnarray}

By projecting the renormalized three-quark operator onto $\Phi_3$ using Eq.(3) of main text,
one readily reproduce the ERBL evolution kernel of $\Phi_3(x)$, whose
explicit expression can be found in Eq.~(4.18) of \cite{Passek-Kumericki:2008uqr}.

\subsection{Analytical expressions of low-lying ${\cal H}^{ijkl}$}

In this section, we enumerate the analytical expressions of various short-distance coefficients
appearing in the nucleon form factor predictions:
\bqa
Q^4 F_1(Q^2) = f_{\rm N}^2(\mu_F) {\sum_{ijkl}} a_{ij}(\mu_F) {\cal H}^{ijkl} a_{kl}(\mu_F),
\label{Master:formula:nucleon:EM:FF}
\eqa

Through NLO accuracy, we have
\begin{eqnarray}
{\cal H}^{ijkl} &=& \frac{1600 \pi^2 \alpha_s^2}{3} \Big\{ e_u\, c^{ijkl}_0 \left[ 1+ {\alpha_s\over \pi}\left( c^{ijkl}_1 L_\mu + c^{ijkl}_2 \right) \right]
 +  e_d\,d^{ijkl}_0 \left[1+ {\alpha_s\over \pi} (d^{ijkl}_1 L_\mu+d^{ijkl}_2) \right] + {\cal O}(\alpha_s^2) \Big\}.
\end{eqnarray}

The various entries read
\begin{itemize}
\item ${\cal H}^{0000}$
\begin{eqnarray}
&& c^{0000}_0 = 1, \; d^{0000}_0 = 2, \; c^{0000}_1 = d^{0000}_1 = \frac{29}{6}, \nn\\
&& c_2^{0000} = -92 \zeta_5+\frac{3719}{30}\zeta_3-\frac{1937}{60},
\; d_2^{0000} = \frac{3}{2}\zeta_5+\frac{209}{120}\zeta_3+\frac{1661}{120}.
\end{eqnarray}

\item ${\cal H}^{1000}$
\begin{eqnarray}
c^{1000}_0 =\frac{49}{6}, \; d^{1000}_0 =- \frac{49}{6}, \; c^{1000}_1 = d^{1000}_1 = \frac{97}{18},
\; c_2^{1000} = d_2^{1000} = -100 \zeta_5+\frac{3203}{40}\zeta_3+\frac{961189}{30240}.
\end{eqnarray}

\item ${\cal H}^{1010}$
\begin{eqnarray}
&& c_0^{1010} = \frac{1519}{9}, \; d_0^{1010} = \frac{1568}{9}, \; c^{1010}_1 = d^{1010}_1 = \frac{107}{18}, \nn\\
&& c_2^{1010} = -\frac{55926 \zeta_5}{217}+\frac{151409 \zeta_3}{868}+\frac{11541697}{133920}, 
\; d_2^{1010} = -\frac{22419 \zeta_5}{224}+\frac{160177 \zeta_3}{2240}+\frac{5573321}{120960}.
\end{eqnarray}

\item ${\cal H}^{1100}$
\begin{eqnarray}
&& c_0^{1100} = \frac{7}{2}, \; d_0^{1100} = -\frac{7}{2}, \; c^{1100}_1 = d^{1100}_1 = \frac{11}{2}, \nn\\
&& c_2^{1100} = -\frac{130}{3}\zeta_5+\frac{5161}{72}\zeta_3-\frac{401}{32},
\; d_2^{1100} = \frac{92}{3}\zeta_5-\frac{11953}{360}\zeta_3+\frac{37937}{1440}.
\end{eqnarray}

\item ${\cal H}^{1110}$
\begin{eqnarray}
c_0^{1110} = \frac{343}{9}, \; d_0^{1110} = -\frac{343}{9}, \; c^{1110}_1 = d^{1110}_1 = \frac{109}{18},
\; c_2^{1110} = d_2^{1110} = -\frac{2400 \zeta_5}{7}+\frac{447161 \zeta_3}{1960}+\frac{23482099}{211680}. 
\end{eqnarray}

\item ${\cal H}^{1111}$
\begin{eqnarray}
&& c_0^{1111} = \frac{539}{9}, \; d_0^{1111} = \frac{196}{9}, \; c^{1111}_1 = d^{1111}_1 = \frac{37}{6}, \nn\\
&& c_2^{1111} = \frac{762}{11}\zeta_5-\frac{19183}{770}\zeta_3-\frac{4945849}{332640},
\; d_2^{1111} = -\frac{141}{4}\zeta_5+\frac{17807}{560}\zeta_3+\frac{737371}{30240} .
\end{eqnarray}

\item ${\cal H}^{2000}$
\begin{eqnarray}
&& c_0^{2000} = \frac{427}{20}, \; d_0^{2000} = \frac{161}{20}, \; c^{2000}_1 = d^{2000}_1 = \frac{103}{18}, \nn\\
&& c_2^{2000} = -\frac{74566}{427}\zeta_5+\frac{18937349}{153720}\zeta_3+\frac{1791733}{30240}, 
\; d_2^{2000} = \frac{1780}{161}\zeta_5-\frac{244199}{57960}\zeta_3+\frac{12557081}{695520}.
\end{eqnarray}

\item ${\cal H}^{2010}$
\begin{eqnarray}
c_0^{2010} = \frac{539}{6}, \; d_0^{2010} = -\frac{539}{6}, \; c^{2010}_1 = d^{2010}_1 = \frac{113}{18}, \; c_2^{2010} = d_2^{2010} = -\frac{119124}{385}\zeta_5+\frac{1590773}{7700}\zeta_3+\frac{23033831}{221760}.
\end{eqnarray}

\item ${\cal H}^{2011}$
\begin{eqnarray}
&& c_0^{2011} = \frac{931}{10}, \; d_0^{2011} = \frac{49}{5}, \; c^{2011}_1 = d^{2011}_1 = \frac{115}{18}, \nn\\
&& c_2^{2011} = -\frac{11792}{19}\zeta_5+\frac{19751197}{47880}\zeta_3+\frac{619980913}{3447360}, 
\; d_2^{2011} = -\frac{7471}{14}\zeta_5+\frac{367657}{1008}\zeta_3+\frac{55349039}{362880}.
\end{eqnarray}

\item ${\cal H}^{2020}$
\begin{eqnarray}
&& c_0^{2020} = \frac{23471}{100}, \; d_0^{2020} = \frac{1372}{25}, \; c^{2020}_1 = d^{2020}_1 = \frac{119}{18}, \nn\\
&& c_2^{2020} = -\frac{3537990}{3353}\zeta_5+\frac{46755193}{67060}\zeta_3+\frac{701158249}{2414160}, 
\; d_2^{2020} = \frac{430209}{784}\zeta_5-\frac{2824781}{7840}\zeta_3-\frac{16515077}{161280}.
\end{eqnarray}

\item ${\cal H}^{2100}$
\begin{eqnarray}
c_0^{2100} = \frac{35}{4}, \; d_0^{2100} = -\frac{35}{4}, \; c^{2100}_1 = d^{2100}_1 = \frac{107}{18}, 
\; c_2^{2100} = d_2^{2100} = \frac{2732}{35}\zeta_5-\frac{179843 }{4200}\zeta_3-\frac{92179}{50400}.
\end{eqnarray}

\item ${\cal H}^{2110}$
\begin{eqnarray}
&& c_0^{2110} = d_0^{2110} = \frac{637}{4}, \; c^{2110}_1 = d^{2110}_1 = \frac{13}{2}, \nn\\
&& c_2^{2110} = \frac{4908}{13}\zeta_5-\frac{447763}{1820}\zeta_3-\frac{49524911}{786240}, 
\; d_2^{2110} = -\frac{9546}{91}\zeta_5+\frac{129473}{1820}\zeta_3+\frac{6308573}{112320}.
\end{eqnarray}

\item ${\cal H}^{2111}$
\begin{eqnarray}
\; c_0^{2111} = \frac{539}{12}, \; d_0^{2111} = -\frac{539}{12}, \; c^{2111}_1 = d^{2111}_1 = \frac{119}{18}, 
\; c_2^{2111} = d_2^{2111} = \frac{47940}{77}\zeta_5-\frac{307441}{770}\zeta_3-\frac{88449677}{665280}.
\end{eqnarray}

\item ${\cal H}^{2120}$
\begin{eqnarray}
c_0^{2120} = \frac{3381}{40}, \; d_0^{2120} = -\frac{3381}{40}, \; c^{2120}_1 = d^{2120}_1 = \frac{41}{6}, 
\; c_2^{2120} = d_2^{2120} = \frac{36348}{161}\zeta_5-\frac{1422557}{9660}\zeta_3-\frac{88420259}{4173120}.
\end{eqnarray}

\item ${\cal H}^{2121}$
\begin{eqnarray}
&& c_0^{2121} = d_0^{2121} = \frac{637}{4}, \; c^{2121}_1 = d^{2121}_1 = \frac{127}{18}, \nn\\
&& c_2^{2121} = -\frac{94014}{91}\zeta_5+\frac{124647}{182}\zeta_3+\frac{673746511}{2358720},
\; d_2^{2121} = \frac{57285}{91}\zeta_5-\frac{106337}{260}\zeta_3-\frac{2490181}{20160}.
\end{eqnarray}

\item ${\cal H}^{2200}$
\begin{eqnarray}
&& c_0^{2200} = \frac{16}{5}, \; d_0^{2200} = \frac{1}{10}, \; c^{2200}_1 = d^{2200}_1 = 6, \nn\\
&& c_2^{2200} = -\frac{3543}{16}\zeta_5+\frac{1924849}{11520}\zeta_3+\frac{7761793}{138240},
\; d_2^{2200} = 840 \zeta_5-\frac{144079}{360}\zeta_3-\frac{1488889}{4320}.
\end{eqnarray}

\item ${\cal H}^{2210}$
\begin{eqnarray}
c_0^{2210} = \frac{49}{4}, \; d_0^{2210} = -\frac{49}{4}, \; c^{2210}_1 = d^{2210}_1 = \frac{59}{9}, 
\; c_2^{2210} = d_2^{2210} = \frac{34068}{35}\zeta_5-\frac{109994}{175}\zeta_3-\frac{4478633}{20160}.
\end{eqnarray}

\item ${\cal H}^{2211}$
\begin{eqnarray}
&& c_0^{2211} = \frac{427}{60}, \; d_0^{2211} = \frac{161}{60}, \; c^{2211}_1 = d^{2211}_1 = \frac{20}{3}, \nn\\
&& c_2^{2211} = -\frac{180108}{61}\zeta_5+\frac{1752946}{915}\zeta_3+\frac{46735883}{58560},
\; d_2^{2211} = -\frac{8352}{23}\zeta_5+\frac{15283}{69}\zeta_3+\frac{3215629}{22080}.
\end{eqnarray}

\item ${\cal H}^{2220}$
\begin{eqnarray}
&& c_0^{2220} = \frac{6069}{200}, \; d_0^{2220} = \frac{987}{200}, \; c^{2220}_1 = d^{2220}_1 = \frac{62}{9}, \nn\\
&& c_2^{2220} = \frac{4267860}{2023}\zeta_5-\frac{28050761}{20230}\zeta_3-\frac{25422835331}{52436160}, 
\; d_2^{2220} = \frac{1066224}{329}\zeta_5-\frac{7005001}{3290}\zeta_3-\frac{6517209757}{8527680}.
\end{eqnarray}

\item ${\cal H}^{2221}$
\begin{eqnarray}
\hspace{-0.25cm}
c_0^{2221} = \frac{287}{40}, \; d_0^{2221} = -\frac{287}{40}, \; c^{2221}_1 = d^{2221}_1 = \frac{64}{9}, 
\; c_2^{2221} = d_2^{2221} = -\frac{2540124}{287}\zeta_5+\frac{16718021}{2870}\zeta_3+\frac{5490319259}{2479680}. 
\end{eqnarray}

\item ${\cal H}^{2222}$
\begin{eqnarray}
&& c_0^{2222} = \frac{779}{100}, \; d_0^{2222} = \frac{43}{25}, \; c^{2222}_1 = d^{2222}_1 = \frac{43}{6}, \nn\\
&& c_2^{2222} = \frac{3727080}{779}\zeta_5-\frac{12158634}{3895}\zeta_3-\frac{11850532139}{10095840}, 
\; d_2^{2222} =  -\frac{20277}{43}\zeta_5+\frac{571137}{1720}\zeta_3+\frac{278183957}{2229120}.
\end{eqnarray}

\end{itemize}

\end{widetext}


\end{document}